\documentclass[epj,final,nopacs]{svjour}
\usepackage{epsf}
\begin{document}
\title{Radiative Effects in the Processes of Hadron Electroproduction}
\author{I.Akushevich, N.Shumeiko, A.Soroko}
\institute{National Center of Particle and High Energy Physics,
220040 Minsk, Belarus}
\date{Received: date / Revised version: date}
\abstract{
An approach to calculate radiative corrections to unpolarized
cross section of
semi-inclusive
electroproduction is developed. An explicit
formulae for the lowest order  QED radiative correction are
presented. Detailed numerical analysis is
performed for the kinematics of experiments at the fixed targets.}

\maketitle

\section{Introduction}
\label{intro}

Semi-inclusive processes of hadron electroproduction have been
recognized long ago \cite{Poli} as an important tool for testing
QCD predictions of nucleon structure because them admit to get
information about quark distributions in the nucleon for each
flavour separately.  Precise analysis of the hadron structure
functions extracted from the experimental data requires, however,
an iterative procedure involving radiative correction (RC) of
these data.  RC to cross section of
semi-inclusive processes can be calculated on basis of Bardin and
Shumeiko covariant approach offered in ref.\cite{BSh} originally
for elastic scattering.  In ref.\cite{ShS} the method was
developed on semi-inclusive processes, where analytical formulae
for the lowest order RC for coincident processes
in electroproduction were found.  In this paper we present
analogous formulae, but in contrary to results of the
ref.\cite{ShS} we do not assume integration  over
hadronic kinematical variables $p_t^2$ and $\phi^h$.  It allows to
calculate the model-independent RC relying only
on the common representation for the hadronic tensor.

Last years the cross sections of the hadron electroproduction
on fixed targets were
measured as functions of azimuthal angles and transversal momentum of
registered particles (see \cite{EMC} and references therein).
However this information is not sufficient to
extract all structure functions involved in the hadronic tensor in wide
kinematical region required for RC calculation.
Therefore we have to use some model
for the semi-inclusive structure functions. Such a model should
give a good initial
approximation for the iterative procedure, and we hope that
the model for structure functions, which can be constructed on
the basis of results
of Mulders and Tangerman \cite{Muld1}, is an appropriate one. It should be
noted that unpolarized   structure functions were considered in
\cite{Muld1} along with the  spin dependent distributions. However, in this
paper we restrict ourselves only by calculation of
RC to unpolarized cross section, and the
consideration of RC to observable quantities
in polarization experiments on hadron electroproduction will be a
subject of a separate publication. Notice that
 for azimuthal
effects we use additionally a model given in ref.\cite{Khoze} (see
also
\cite{Karo}).

In the Section \ref{Sekin} we shortly describe the kinematics of
the hadron electroproduction process with and without radiation of
additional photon. The hadronic tensor and the model for structure
functions are discussed in Section \ref{Sebor}. The analytical
formulae for Born cross section and for RC of
the lowest order are given in Sections \ref{Sebor} and
\ref{Serad}.
In this paper the only ultrarelativistic approximation is made:
electron mass is considered to be small.
We note that
final analytical formulae are written
in the form similar to used in FORTRAN code POLRAD 2.0
\cite{POLRAD20}. The Section \ref{Senum} is devoted to numerical
analysis,
 performed on the basis of new code HAPRAD specially developed by us
for this purpose. Most cumbersome formulae are gathered in the
Appendix.

\section{The kinematics} \label{Sekin}
The cross section of hadron $h$  electroproduction
\begin{equation}
 e(k_1) + N (p) \longrightarrow e'(k_2) + h(p_h) + X(p_x)
\end{equation}
depends on five kinematical variables which can be chosen as
\begin{equation}
 x,\;y;\,z,\;t,\;\phi_h,
\label{002}\end{equation}
where $x$ and $y$ are usual scaling variables, $z$ and $t$ are
defined via hadron momentum
\begin{equation}
t=(q-p_h)^2, \quad z=p_hp/pq, \quad q=k_1-k_2,
\label{005}\end{equation}
$\phi_h$ is an angle between planes (${\vec k_1},{\vec k_2}$) and
(${\vec q},{\vec p_h}$) in the rest frame
($p=(M,{\vec 0})$).
 Also the following invariants will be used
\begin{eqnarray}
\label{invar}
&&S = 2k_1p,\; X = 2k_2p = (1-y)S,\; Q^2 = -q^2=xyS,
\nonumber\\&&
W^2=S_x-Q^2+M^2,\;S_{x}= S-X,\;S_p=S+X,\;
\nonumber\\&&
\lambda_Q=S_x^2+4M^2Q^2,
\nonumber\\&&
 M_x^2=p_x^2=(1-z)S_x+M^2+t,\;
V_{1,2}=2k_{1,2}p_h.
\end{eqnarray}

When the radiative process
\begin{equation}
 e(k_1) + N (p) \longrightarrow e'(k_2)+\gamma(k) + h(p_h) +
X({\tilde p_x})
\label{001r}\end{equation}
is considered, the three additional independent variables have to
be introduced
\begin{equation}
 R=2kp, \quad \tau=qk/kp,\quad \phi_k,
\label{002a}\end{equation}
$\phi_k$ is the
rest frame
 angle between planes (${\vec k_1},{\vec k_2}$) and
(${\vec q},{\vec k}$). Also we introduce the
quantity $\mu = kp_h/kp$ and the following invariants
\begin{eqnarray}\label{008}
 {\tilde Q^2}  &=&  (q-k)^2  =  Q^2+R\tau, \nonumber\\
 {\tilde W^2}  &=&  (p+q-k)^2  =  W^2-R(1+\tau), \nonumber\\
 {\tilde t}  &=&  (q-k-p_h)^2  = t+R(\tau-\mu), \nonumber\\
 {\tilde M_x^2}  &=&  {\tilde p_x}^2  = M_x^2+R(1+\tau-\mu).
\end{eqnarray}

The phase space of the three final particles is parameterized as
\begin{equation}
{d^3{\vec k_2} \over k_{20} }
{d^3{\vec p_h} \over p_{h0} }
{d^3{\vec k} \over k_{0} }
=\pi S_x dxdy \; {S_xdzdtd\phi_h \over 2\sqrt{\lambda_Q}}
\; {RdRd\tau d\phi_k \over \sqrt{\lambda_Q}}.
\label{011}\end{equation}
Instead of $t-$dependence we will also consider the cross section as a
function of transversal momentum $p_t$ defined in (\ref{a070}).

We are interesting in explicit dependence on angles $\phi_h$ and
$\phi_k$. So it is useful to take the some scalar  products with
$p_h$ in the form
\begin{eqnarray}\label{009}
 &&   \frac{1}{2}V_{1,2}=k_{1,2}p_h =  a^{1,2}+b\cos\phi_h,
 \\
 &&  \frac{1}{2}\mu R= kp_h  =
R(a^{k}+b^k(\cos\phi_h\cos\phi_k+\sin\phi_h\sin\phi_k)).
\nonumber\end{eqnarray}
Also we will use $a^{\pm}=a^2\pm a^1$. The explicit expressions for coefficients
are given in Appendix (\ref{a065}).

\section{The hadronic tensor and the Born approximation}
\label{Sebor}

The cross section of the electroproduction process can be obtained
in terms of convolution of leptonic and hadronic tensors. There
are two leptonic tensors: with and without additional radiated
photon. The Born leptonic tensor (without a photon) is standard,
and radiative one is cumbersome. The explicit expressions for the
leptonic tensors and formulae for RC in terms of them can be found
in
\cite{AISh,Ak}.

The hadronic tensor without the $T$- and $P$-odd terms can be
presented in the form \cite{hagiwara,pak}
\begin{equation}
W^{\mu\nu}=-{\tilde g}^{\mu\nu}{\cal H}_1
+{\tilde p}^{\mu}{\tilde p}^{\nu}{\cal H}_2
+{\tilde p_h}^{\mu}{\tilde p_h}^{\nu}{\cal H}_3
+( {\tilde p}^{\mu}{\tilde p_h}^{\nu}
  +{\tilde p_h}^{\mu}{\tilde p}^{\nu}
     ){\cal H}_{4},
\label{003}\end{equation}
where
\begin{equation}
{\tilde g}^{\mu\nu}=g^{\mu\nu}+\frac{q^{\mu}q^{\nu}}{Q^2}
,\quad
{\tilde p}^{\mu}=p^{\mu}+\frac{q^{\mu}\;pq}{Q^2}
,\quad
{\tilde p_h}^{\mu}=p_h^{\mu}+\frac{q^{\mu}\;p_hq}{Q^2},
\label{004}\end{equation}
and all of the SF depend on four kinematical invariants
(for example, $Q^2$, $W^2$, $t$, $z$)
The model for structure functions can be constructed on basis of results
of the paper Mulders and Tangerman \cite{Muld1}. Keeping only
the leading twist contribution  we have for structure functions
\begin{eqnarray} \label{model}
{\cal H}_{1} &=& \sum_q e_q^2 f_q(x)D_q {\cal G}   
,\nonumber\\[0.3cm]
{\cal H}_{2} &=& -\frac{p_t^2+m_h^2}{M^2E_h^2}\sum_q e_q^2 f_q(x)D_q {\cal  G}
,\nonumber\\[0.3cm]
{\cal H}_{3} &=& 0
,\nonumber\\[0.3cm]
{\cal H}_{4} &=& \frac{1}{ME_h}\sum_q e_q^2 f_q(x)D_q {\cal G},
\end{eqnarray}
and
\begin{equation}
{\cal G}={\cal G}_1=b\exp(-bp_t^2)
,
\label{calG}\end{equation}
where $b=R^2/z^2$ is a slope parameter and $R$ is a parameter of the model.

The Born cross section has the following dependence on $\phi_h$:
\begin{equation}
\sigma_0 = {d\sigma_0\over dxdydzdp_t^2d\phi_h}=
\frac{N}{Q^4}
(A+\cos\phi_h A^c+\cos
2\phi_h A^{cc}),
\label{010}\end{equation}
where $N=\alpha^2 y S_x/\sqrt{\lambda_Q}$.
The coefficients $A$ do not depend on $\phi_h$ more and they have
the form
\begin{eqnarray}
A&=&
 2Q^2{\cal H}_1
+ (SX - M^2Q^2){\cal H}_{2}
\nonumber\\&&
+ ( 4a^1a^2 + 2b^2-M_h^2Q^2){\cal H}_{3}
\nonumber\\&&
+ (2Xa^1+2Sa^2 - zS_xQ^2){\cal H}_{4},
\nonumber\\
A^c&=& 2b(2a^+{\cal H}_{3} + S_p{\cal H}_{4}),
\nonumber\\
A^{cc}&=& 2b^2{\cal H}_{3}.
\end{eqnarray}
When integrated over the kinematical variables $\phi_h$ and $p_t$ the
cross section (\ref{010}) coincides with the well known formula
for semi-inclusive cross section calculated within QPM
\begin{equation}
 \sigma_{xyz}={d\sigma \over dxdydz}={2\pi\alpha^2 \over y Q^2}
(y^2+2-2y) \sum_q e_q^2 f_q(x)D_q.
\end{equation}
Unfortunately, the $p_t^2$ distribution
\begin{equation}
{1\over \sigma_{xyz}}
{d \sigma_{xyz} \over p_t^2}
\approx {\cal G},
\end{equation}
calculated with the exponential slope (\ref{calG}) does not fit
experimental
data with sufficient $\chi^2$. So the more complicated model \cite{JM}
with power dependence on $p_t^2$ seems to be more adequate. Another
possibility to come to an agreement with the data consists in replacement
of the Gaussian factor (\ref{calG}) by the fit of experimental
$p_t^2$-distribution taken in the form \cite{EMC}
\begin{equation}
{\cal G}={\cal G}_2=\biggl[{1\over a+bz+p_t^2}\biggr]^{c+dz}.
\label{calG2}\end{equation}

\begin{table*}[!t]
\begin{center}
\begin{tabular}{c|c|c|c|c|c|c|c|c|c|c}\hline
&&&&SIRAD&\multicolumn{6}{c}{HAPRAD}\\ \cline{5-11}
$x$ & $y$ & $Q^2$  & $z$ &  &  \multicolumn{3}{c|}{without  cuts}&
  \multicolumn{3}{c}{with cuts}
 \\ \cline{6-11}
 &&GeV$^2$&&&${\cal G}_1$&${\cal G}'_1$&${\cal G}_2$&${\cal G}_1$
 &${\cal G}'_1$&${\cal G}_2$
\\
\hline
  \hline
0.038 & 0.677 & 1.33 & 0.25 & 1.029 &1.033&1.024&0.982& 1.041&1.025&0.985\\ 
0.062 & 0.567 & 1.82 & 0.35 & 0.996 &0.989&0.989&0.947& 0.989&0.980&0.951\\ 
0.092 & 0.529 & 2.52 & 0.45 & 0.970 &0.961&0.961&0.934& 0.961&0.956&0.936\\ 
0.131 & 0.499 & 3.38 & 0.55 & 0.945 &0.936&0.933&0.912& 0.934&0.931&0.906\\ 
0.198 & 0.476 & 4.88 & 0.65 & 0.918 &0.902&0.902&0.889& 0.897&0.897&0.881\\ \hline
\end{tabular}
\end{center}
\caption{The results for RC factors to three dimensional semi-inclusive
cross section obtained using FORTRAN codes SIRAD and HAPRAD (see
text for further explanations). Kinematical points are taken from the
Table 1 of ref.\cite{Manuella}.
}
\label{tab1}
\end{table*}

\section{The radiative correction of the lowest order} \label{Serad}

The cross section that  takes into account radiative effects can
be written as
\begin {equation}
\sigma _{obs} = \sigma _0 e^{\delta_{inf}}
(1+ \delta_{VR}+\delta_{vac})+\sigma_{F}.
\label{eq1}
\end {equation}
Here the corrections
 $\delta_{inf}$  and   $\delta_{vac}$  come from  radiation of soft
photons \cite{Sh} and effects of vacuum polarization\footnote{There are explicit
formulae for leptonic contribution to vacuum polarization effect
(see \protect\cite{ASh} for example) and parameterization of hadronic one
\protect\cite{delvac}.}.
The correction  $\delta_{VR}$ is infrared free sum of factorized
parts of real and virtual photon radiation. These quantities are
given by the following expressions
\begin{eqnarray}        \label{deltas}
\delta_{VR} &=&\frac{\alpha}{\pi}
\biggl(\frac{3}{2}l_m\!-\!2\!-\frac{1}{2}\ln^2\frac{X'}{S'}+{\rm
Li}_2
\frac{S'X'-Q^2p_x^2}{S'X'}-\frac{\pi^2}6\biggr)\!,
\nonumber\\
\delta _{inf}&=& \frac{\alpha}{\pi}
(l_m-1)\ln\frac{(p_x^2-(M+m_{\pi})^2)^2}{S'X'},
\\
\delta _{vac}&=& \delta_{vac}^{lept}+\delta_{vac}^{hadr},\nonumber
\end{eqnarray}
where $S'=X+Q^2-V_2$, $X'=S-Q^2-V_1$, $l_m=\ln Q^2/m^2$ and ${\rm
Li}_2$
is Spence function or dilogarithm.

The contribution of radiative tail has the standard form
\cite{ASh,Ak}
\begin{eqnarray}\label{020}
\sigma _F &=& -{\alpha N \over 2 \pi}
 \int\limits^{2\pi }_{0}d\phi_k
 \int\limits^{\tau_{max}}_{\tau_{min}}d\tau
 \sum^{4}_{i=1}
 \sum^{3}_{j=1} \theta
_{ij}(\tau,\phi_k) \times
\nonumber \\ && \qquad \times
\int\limits^{R_{max}}_{0}
dR R^{j-2} \biggl[
{{\cal H}_i\over (Q^2+R\tau)^{2}}-
\delta_j{{\cal H}_i^0\over Q^4}\biggr] .
\end{eqnarray}
Here $2M^2\tau_{max,min}=S_x\pm \sqrt{\lambda_Q}$ and $R_{max}
=(M_x^2-(M+m_{\pi})^2)/(1+\tau-\mu)$, $\delta_j$=1 for
$j$=1 and $\delta_j$=0 otherwise. The explicit formulae for
functions
 $\theta(\tau,\phi_k)$ can be found in Appendix.
The structure functions ${\cal H}_{i}^0 $ have to be calculated for Born
kinematics, but ${\cal H}_{i}$ is calculated in terms of tilde
variables (\ref{008}).

\section{Numerical analysis} \label{Senum}

In this section we give numerical results for RC to semi-inclusive
unpolarized cross section. For all cases RC factor is defined as a ratio of
observed to born cross sections. Also we will speak about relative RC (or
simply RC), which is difference between observed and born cross sections or
asymmetries divided on Born ones.
 
For definiteness we choose the
kinematics of experiment HERMES at DESY. First, HERMES is the
modern current experiment with rich possibilities for studies
in semi-inclusive physics (see \cite{Manuella}). Second, HERMES is
the experiment with electron beam, so the relatively large RC
in respect to muon DIS experiment are anticipated.

The dependence of RC on $x$, $y$ and $z$ is widely discussed in
ref.\cite{ShS}, where quark-parton model was assumed. Similar
results can be obtained using
the formulae of this paper after integration over $p_t$ and $\theta_h$.
In this paper we
concentrate on comparison of the codes constructed on basis of the
two sets of formulae and on studying of the effects beyond the
quark-parton model: azimuthal asymmetries and dependencies on
transversal momentum $p_t$.

\subsection{FORTRAN codes POLRAD 2.0 and HAPRAD}
\label{codes}

The special FORTRAN code HAPRAD was developed to calculate the RC
to five-dimensional cross section\penalty -10000
$d^5\sigma/dxdydzdp_t^2d\theta_h$.~From~ the other hand there
is the code POLRAD 2.0 with special patch SIRAD, which calculated
RC to semi-inclusive three dimensional cross section obtained in
QPM. 

In this section we show that numerical results for RC
to $d\sigma/dxdydz$
reproduced by these two codes coincide with a good accuracy. It
can be seen from the Table \ref{tab1}. In the table we represent
RC to the cross
section as it follows from the runs of the codes POLRAD 2.0 and HAPRAD.
Since HAPRAD allows one to take into account the kinematical cuts
and to
use different models for $p_t^2$-slope,
three fits for $p_t^2$-distribution
and
the cases with and without experimental cuts were considered.  The
first fit for $p_t^2$-slope is defined in (\ref{calG}),  while
the second and third ones are our
fits of experimental data \cite{EMC} using exponential
(${\cal G}'_1={\cal G}_1$ at $b=a/z$) and power (\ref{calG2}) functional
forms.
 As the kinematical
cuts on
$\phi_h$ and $p_t$ of the measured hadron we took HERMES geometrical
ones \cite{HERMES}.
 We can conclude from this analysis that
 neither important differences between SIRAD and HAPRAD results,
if exponential model for $p_t^2$-distribution is used,   nor
dependence on
slope parameter model and applying of geometrical cuts are found.
However RC takes a negative shift in the case of the model
based on power functional form (\ref{calG2}). As it is discussed
below RC depends on steepness of $p_t^2$ distribution. It is a
reason why models like $\delta(p_t^2)$ (QPM, POLRAD 2.0) and
(\ref{calG}) give larger RC. Within practical RC procedure
in concrete measurement of $d\sigma/dxdydz$ the model can be
fixed only if the information about $p_t^2$-distribution is
considered additionally.

Also two models for fragmentation function were considered: simple
parameterization of the pion data \cite{Aubert} and modern model
in the next-to-leading order QCD \cite{Kramer}. RC factor
calculated using these models differs
on several per cent. However this model dependence is less
important because it can be eliminated by applying an iteration
procedure, where fit of extracted data is used for RC calculation
in subsequent step.

\begin{figure}[t]
\unitlength 1mm
\begin{picture}(80,80)
\put(33,69){\makebox(0,0){\small $y=0.2$}}
\put(72,69){\makebox(0,0){\small $y=0.4$}}
\put(33,33){\makebox(0,0){\small $y=0.6$}}
\put(72,33){\makebox(0,0){\small $y=0.8$}}
\put(10,38){\makebox(0,0){\small $\delta$}}
\put(10,75){\makebox(0,0){\small $\delta$}}
\put(50,38){\makebox(0,0){\small $\delta$}}
\put(50,75){\makebox(0,0){\small $\delta$}}
\put(40,40){\makebox(0,0){\small $z$}}
\put(80,40){\makebox(0,0){\small $z$}}
\put(40,3){\makebox(0,0){\small $z$}}
\put(80,3){\makebox(0,0){\small $z$}}
\put(0,-5){
\epsfxsize=9cm
\epsfysize=9cm
\epsfbox{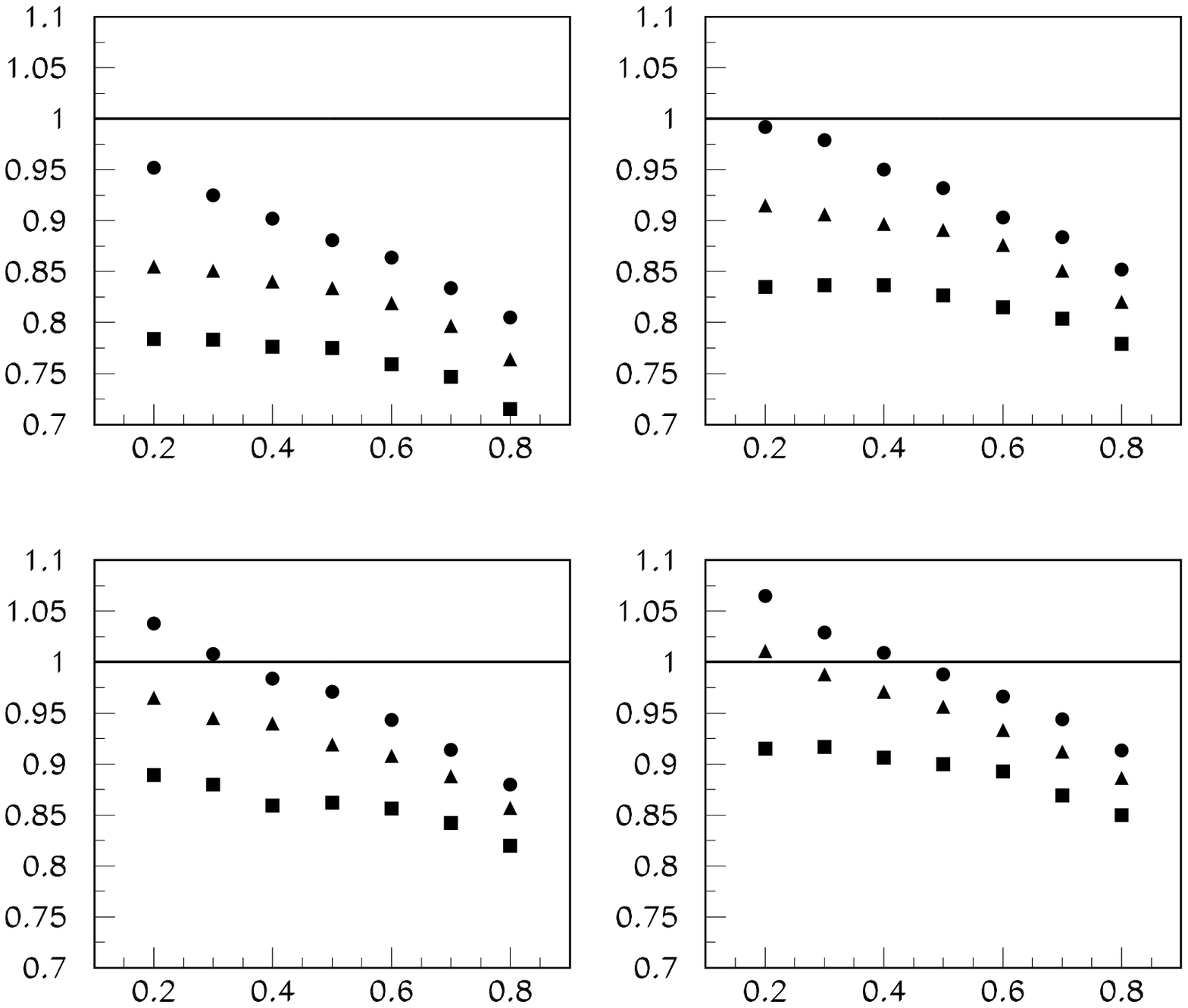}
}
\end{picture}
\caption{\protect\it Radiative correction to the semi-inclusive cross
section for kinematics of HERMES; $\sqrt{S}$=7.19 GeV. Symbols
from top to bottom correspond to the
$x$=0.05,
0.45 and
0.7. The results for $x$=0.15 are skipped, because they practically
coincide with ones for $x$=0.05.
}
\label{del}
\end{figure}

\subsection{Cross section and $<\!\! p_t^2 \!\!>$}

\begin{figure}[t]
\unitlength 1mm
\begin{picture}(80,80)
\put(65,7){\makebox(0,0){$p_t/p_{t\;max}$}}
\put(13,77){\makebox(0,0){ $\bar\delta$}}
\put(15,70){\makebox(0,0){\small $z=0.2$}}
\put(41,67){\makebox(0,0){\small $z=0.4$}}
\put(61,65){\makebox(0,0){\small $z=0.6$}}
\put(66,49){\makebox(0,0){\small $z=0.2$}}
\put(66,45){\makebox(0,0){\small $z=0.4$}}
\put(66,38){\makebox(0,0){\small $z=0.6$}}
\put(66,27){\makebox(0,0){\small $z=0.8$}}
\put(0,5){
\epsfxsize=8cm
\epsfysize=8cm
\epsfbox{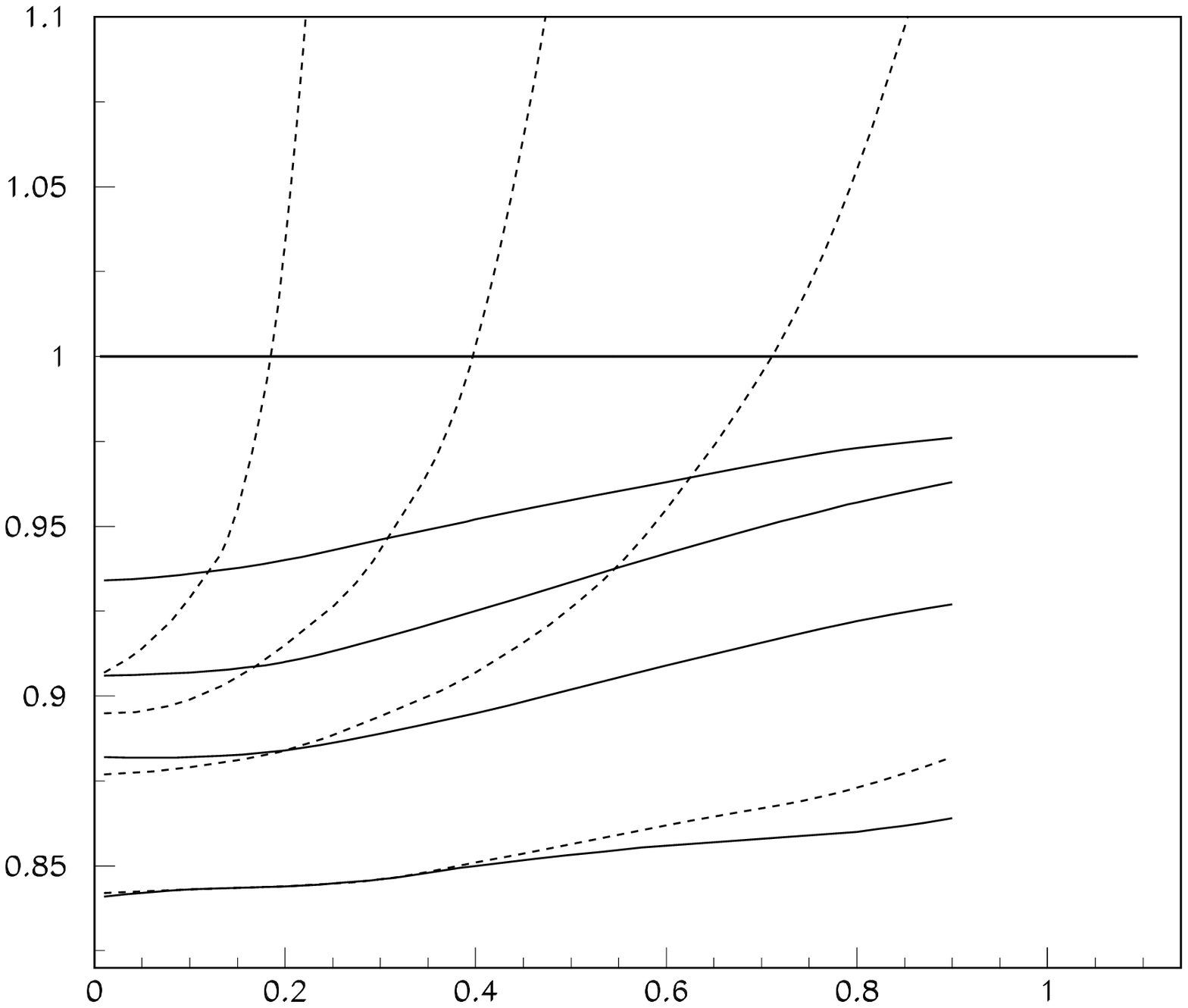}
}
\end{picture}
\caption{\protect\it Radiative correction to semi-inclusive cross section
vs $p_t$; $\sqrt{S}$=7.19 GeV, $x$=0.15, $Q^2$=4 GeV$^2$. Dashed and solid
curves correspond to models (\ref{calG}) and (\ref{calG2})
respectively.
}
\label{ptpt}
\end{figure}

Here we give numerical results for unpolarized cross section in
kinematics of experiment HERMES \cite{HERMES}.
RC factor ($\delta$) to the semi-inclusive cross section integrated over
$p_t$ and $\phi_h$ as a function of $x$, $y$ and $z$ is presented in 
Figure \ref{del}.
Further we
analyse the $z$ and $p_t$ dependence of the cross section and
azimuthal asymmetries.
The dependencies of RC factor
($\bar\delta=\bar\sigma_{obs}/\bar\sigma_{born}$) to the semi-inclusive
cross section  on hadronic
variables
$z$ and $p_t$
are  shown
in Figure \ref{ptpt},
where  sigma bar ($\bar\sigma$) is meant the four-dimensional
 cross section $d\sigma/dxdydzdp_t^2$.
 We note that the obtained large correction to
the cross section vs $p_t$ is an analog of the similar results for vector
meson electroproduction \cite{Ak}. In our case the slope parameter depends
on $z$ (see (\ref{calG})), so we have important $z$-dependence of the
effect.
However, if the experimental fit for $\cal G$ is used (solid curves in
the Figure \ref{ptpt}) there are no such a rise of RC for high
$p_t^2$ values.

\begin{figure}[t]
\unitlength 1mm
\begin{picture}(80,80)
\put(21,45){\oval(6.0,6.0)[rt]}
\put(21,51){\oval(6.0,6.0)[rb]}
\put(26,61){\oval(4.0,4.0)[lt]}
\put(26,35){\oval(4.0,4.0)[lb]}
\put(24,61){\line(0,-2){10.}}
\put(24,45){\line(0,-2){10.}}
\put(69,6){\makebox(0,0){$z$}}
\put(57,62){\makebox(0,0){1) ${\cal G}={\cal G}_1$ (\ref{calG})}}
\put(57,56){\makebox(0,0){2) ${\cal G}={\cal G}_2$ (\ref{calG2})}}
\put(19,48){\makebox(0,0){$1$}}
\put(23,20){\makebox(0,0){$2$}}
\put(20,70){\makebox(0,0){${\bar \delta}_{pt}$}}
\put(5,0){
\epsfxsize=8cm
\epsfysize=8cm
\epsfbox{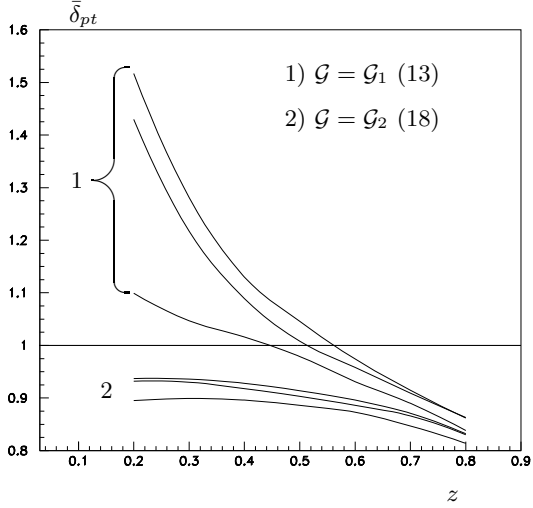}
}
\end{picture}
\caption{\it
Radiative correction  to $<\!\! p_t^2 \!\!>$ 
defined in eq. (\ref{pt02}) for  HERMES kinematics, $\sqrt{S}=$7.19
GeV,
y=0.4. Curves from top to bottom corresponds to $x$=0.15, 0.05 and 0.45.
}
\label{figpt}
\end{figure}

The crucial for QCD predictions \cite{Poli} quantity
 $<\!\! p_t^2 \!\!>$ is expressed in terms of $\bar\sigma$ as
\begin{equation}
<\!\! p_t^2 \!\!>=
{\int dp_t^2 \; p_t^2 {\bar\sigma} \over
\int dp_t^2 \; {\bar\sigma}} .
\label{pt01}\end{equation}
RC to this
quantity
can be expressed as
\begin{eqnarray}\label{pt02}
\delta_{pt}&=&
{\int dp_t^2 \; p_t^2 \;{\bar\sigma_{obs}}
\over \int dp_t^2 \; {\bar\sigma_{obs}}}
\bigg/
{\int dp_t^2 \; p_t^2 \;{\bar\sigma_{born}} \over \int dp_t^2 \;
{\bar\sigma_{born}}}=
\nonumber \\ && =
{\int dp_t^2 \; p_t^2 \;{\bar\sigma_{obs}} \over
\int dp_t^2 \; p_t^2 \;{\bar\sigma_{born}}}
{\int dp_t^2 \; {\bar\sigma_{born}} \over
\int dp_t^2 \; {\bar\sigma_{obs}}}
=
\frac{\bar\delta_{pt}}{\delta},
\end{eqnarray}
where $\delta$ is semi-inclusive RC factor discussed above
(see Figure \ref{del}). The correction $\bar\delta_{pt}$ both for 
exponential and
power model (\ref{calG},\ref{calG2}) is
presented on Figure \ref{figpt}.

Similar to the case of $p_t^2$-distribution the radiative effect is
larger when exponential model (\ref{calG}) is used. That is
because of
contribution of small  $\tilde p_t^2$ when we integrate over the phase
space of emitted photon (\ref{020}). The steeper the slope of the
$p_t^2$-distribution the larger this contribution to RC.

\subsection{Azimuthal asymmetries}

The following azimuthal asymmetries are measurable in the
experiments \cite{EMCazim}
\begin{eqnarray}
<\!\cos \phi_h\!>&=&\bar\sigma^{-1}_{obs}
\int\limits_{0}^{2\pi}d\phi_h \;\cos\phi_h \; \sigma_{obs},
\nonumber\\[0.3cm]
<\!\cos 2\phi_h\!>&=&\bar\sigma^{-1}_{obs}
\int\limits_{0}^{2\pi}d\phi_h
\;\cos2\phi_h \; \sigma_{obs},
\nonumber\\[0.3cm]
<\!\sin \phi_h\!>&=&\bar\sigma^{-1}_{obs}
\int\limits_{0}^{2\pi}d\phi_h
\;\sin\phi_h \; \sigma_{obs}.
\end{eqnarray}
In terms of these quantities the observed cross section
(\ref{eq1}) can be written as
\begin{eqnarray}
\sigma_{obs}&=&\bar\sigma_{obs} \bigl(1+
<\!\cos \phi_h\!> \cos \phi_h +
\nonumber\\ &&
\qquad\qquad 
+<\!\cos 2\phi_h\!> \cos 2\phi_h
\bigr)+
 \sigma_{add}.
\end{eqnarray}
  Here $\sigma_{add}$ is originated from contribution of higher
harmonics ($\sin \phi_h$, $\sin 2\phi_h$, ...). There are no their
contributions at the Born level (see eq.\ref{010}), and
$\sigma_{add}^{Born}=0$.

\begin{figure}[t]
\unitlength 1mm
\begin{picture}(80,80)
\put(69,6){\makebox(0,0){$z$}}
\put(20,70){\makebox(0,0){$<\!\!cos(\phi_h)\!\!>$}}
\put(5,0){
\epsfxsize=8cm
\epsfysize=8cm
\epsfbox{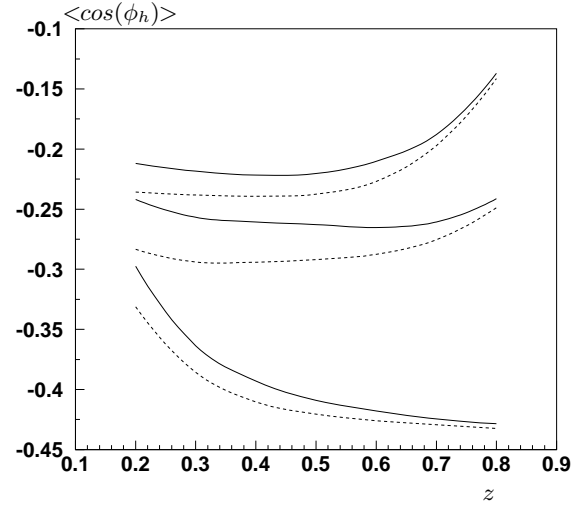}
}
\end{picture}
\caption{\it
Azimuthal asymmetry
$<\protect\!\protect\!\protect\cos \protect\phi_h\protect\!\protect\!>$ vs
$z$ for $y=$0.2 within HERMES kinematics; $\sqrt{S}$=7.19 GeV.
Dashed (solid) lines correspond to born(observed) asymmetries.
Curves from top to bottom correspond to $x$=0.7, 0.45 and 0.05.
}
\label{Fig3}
\end{figure}

Azimuthal asymmetry $<\!\!\cos \phi_h\!\!>$ is negative in considered
region. RC to the quantity can exceed 10\%
The result is done in Fig.\ref{Fig3}.

Within the model (\ref{model}) $<\!\!\cos 2\phi_h\!\!>$
is equal to zero at the Born level. So the asymmetry in this
case is defined by RC only. Our estimation shows that this
effect is of order 1\%. Relative RC to the asymmetry can be
 estimated using another model \cite{Khoze}, where
$<\!\!\cos 2\phi_h\!\!>\neq 0$ at the Born level. It is of order 10\% in the
region of applicability of the model.

$<\!\!\sin \phi_h\!\!>$ is equal to zero at the Born level in any
case. But there is nonzero contribution to it coming from
RC. The numerical analysis shows that radiative
corrections does not give visible contribution to it. For HERMES
kinematical region values of $<\!\!\sin \phi_h\!\!>$ does not exceed
0.01\%.

We should stress that our predictions for the values of the radiative
effects display a strong model dependence. Therefore any reliable
method of radiative correction of experimental data has to be based
on an iterative procedure, where all necessary fits for RC codes use
the processing data as an input and are specified in every step of this
procedure. This procedure can be readily developed on the basis of the
code HAPRAD.

\section{Discussion and Conclusion}

In this paper the QED radiative correction to different observable
quantities in the experiments on hadron electroproduction is
analyzed. The explicit covariant formulae are given in Section
\ref{Serad} and Appendix.

New FORTRAN code HAPRAD
is
developed in
order to perform the numerical analysis. It was shown in Section
\ref{codes} that the results for the RC to cross section
integrated over $p_t$ and $\phi_h$ are in agreement with POLRAD
2.0 \cite{POLRAD20}. Several models for structure functions and
slope parameter in respect to $p_t^2$ were applied for. It is
found that the model based on power $p_t^2$-slope model leads to
smaller values for RC.

Within the exponential model for $\cal G$ (\ref{calG}) RC to $<\!\! p_t^2
\!\!>$ can exceed 40\%. However it
 essentially depends on the model of
$p_t^2$-distribution as well as on $x$ and $z$.

RC to azimuthal asymmetries is of order 10\%.  The asymmetry $<\!\!\sin
\phi_h\!\!>$ due to RC is found to be negligible but not equaling to
zero exactly.

FORTRAN code HAPRAD is available
(aku@hep.by) for the calculation of RC to observable quantities in
the experiments on the hadron electroproduction.

\section*{Acknowledgements}
We are grateful to H.Avakian, A.Brull, H.Ihssen, R.Milner,
K.Oganessyan
and
H.Spiesberger for
fruitful discussions and comments.

\appendix
\renewcommand{\thesection}{Appendix}
\renewcommand{\thesubsection}{\Alph{section}.\arabic{subsection}}
\renewcommand{\theequation}{\Alph{section}.\arabic{equation}}
\section{}
\setcounter{equation}{0}

In this appendix we list the explicit form for functions
$\theta_{ij}$:
\begin{equation}
\theta_{ij}(\tau,\phi_k)=
\theta_{ij}^0+
\cos{\phi_h}\theta_{ij}^c+
\sin{\phi_h}\theta_{ij}^s+
\cos{2\phi_h}\theta_{ij}^{cc},
\label{a01}\end{equation}
where
\begin{eqnarray}
\theta^{0}_{12}&=& 4 \tau  F_{IR}
,\nonumber\\[0.3cm]
\theta^{0}_{22}&=&  - \frac{1}{2} F_{d} S_p^2 \tau + \frac{1}{2} F_{1+} S_p S_x
\nonumber\\&&
+ F_{2-} S_p
+ 2 F_{2+} M^2 \tau
\nonumber\\&&
 - 2 F_{IR} M^2 \tau
 + F_{IR} S_x
,\nonumber\\[0.3cm]
\theta^{0}_{32}&=& 2 ( - 2 F_{d} b^2 \tau -  F_{d} (a^+)^2 \tau -
F_{1+} a^
- a^+
\nonumber\\&&
+ 2 F_{2-} \cos\phi_k b b^k
+ 2 F_{2-} a^k a^+
\nonumber\\&&
-  F_{IR} M_h^2 \tau - 2 F_{IR} a^k a^-)
,\nonumber\\[0.3cm]
\theta^{c}_{32}&=& 4  ( - 2 F_{d} b a^+ \tau -  F_{1+} b a^- +
F_{2-} \cos\phi_k b^k a^+
\nonumber\\&&
+ 2 F_{2-} b a^k - \cos\phi_k F_{IR} b^k a^-)
,\nonumber\\[0.3cm]
\theta^{cc}_{32}&=& 4  b ( -  F_{d} b \tau + F_{2-} \cos\phi_k b^k)
,\nonumber\\[0.3cm]
\theta^{s}_{32}&=& 4 b^k \sin\phi_k  (F_{2-} a^+ - F_{IR} a^-)
,\nonumber\\[0.3cm]
\theta^{0}_{42}&=&  - 2 F_{d} a^+ S_p \tau - F_{1+} a^- S_p +
F_{1+} a^+ S_x
\nonumber\\&&
 + 2F_{2-} a^k S_p + 2F_{2-} a^+
 + 2F_{2+} \tau z S_x
\nonumber\\&&
 + F_{IR} a^k S_x
 - 2F_{IR} a^- - 2 F_{IR} \tau z S_x
,\nonumber\\[0.3cm]
\theta^{c}_{42}&=& 2  ( - 2 F_{d} b S_p \tau +  F_{1+} b S_x +
F_{2-}
\cos\phi_k b^k S_p
\nonumber\\&&
+ 2 F_{2-} b + \cos\phi_k F_{IR} b^k S_x)
,\nonumber\\[0.3cm]
\theta^{s}_{42}&=& 2 b^k \sin\phi_k  (F_{2-} S_p + F_{IR} S_x)
,\nonumber\\[0.3cm]
\theta^{0}_{13}&=&  - 2 (F + F_{d} \tau^2)
,\nonumber\\[0.3cm]
\theta^{0}_{23}&=& 2  F M^2 +  F_{d} M^2 \tau^2 - \frac{1}{2} F_{d} S_x \tau -
\frac{1}{2} F_{1+} S_p
,\nonumber\\[0.3cm]
\theta^{0}_{33}&=& 2  FM_h^2 + F_{d} M_h^2 \tau^2 + 2 F_{d} a^k a^-
\tau
\nonumber\\&&
- 2 F_{1+} \cos\phi_k b b^k - 2 F_{1+} a^k a^+
,\nonumber\\[0.3cm]
\theta^{c}_{33}&=& 2  (F_{d} \cos\phi_k b^k a^- \tau - F_{1+}
\cos\phi_k b^k a^+ -
2 F_{1+} b a^k)
,\nonumber\\[0.3cm]
\theta^{cc}_{33}&=&  - 2 F_{1+}  \cos\phi_k b b^k
,\nonumber\\[0.3cm]
\theta^{s}_{33}&=& 2 b^k \sin\phi_k  (F_{d} a^- \tau - F_{1+} a^+)
,\nonumber\\[0.3cm]
\theta^{0}_{43}&=& 2 F z S_x - F_{d} a^k S_x \tau + F_{d} a^- \tau +
F_{d} \tau^2
z S_x
\nonumber\\&&
- F_{1+} a^k S_p - F_{1+} a^+
,\nonumber\\[0.3cm]
\theta^{c}_{43}&=&  - F_{d} \cos\phi_k b^k S_x \tau - F_{1+}
\cos\phi_k b^k S_p -
2 F_{1+} b
,\nonumber\\[0.3cm]
\theta^{s}_{43}&=&  -  b^k \sin\phi_k  (F_{d} S_x \tau + F_{1+}
S_p)
.\end{eqnarray}
Here
\begin{eqnarray}\label{030}
F_{1+}=\frac{F}{z_1}+\frac{F}{z_2}, &&
F_{2+}=F\biggl(\frac{m^2}{z_2^2}+\frac{m^2}{z_1^2}\biggr),
\nonumber\\[0.3cm]
F_{2-}=F\biggl(\frac{m^2}{z_2^2}-\frac{m^2}{z_1^2}\biggr), &&
F_{d}=\frac{F}{z_1z_2},
\end{eqnarray}
where $F=1/(2\pi\sqrt{\lambda_Q})$
and
\begin{equation}
  F_{IR}=F_{2+}-Q^2F_d.
\label{031}\end{equation}
The quantities
$z_{1,2}=kk_{1,2}/kp$ in terms of integration variables can be expressed as
\begin{eqnarray}\label{a05}
z_1&=&\lambda_Q^{-1}\bigl(Q^2S_p+\tau(SS_x+2M^2Q^2)-2M\cos\phi_k\sqrt{\lambda_{\tau}\lambda}\bigr),
\nonumber\\
z_2&=&\lambda_Q^{-1}\bigl(Q^2S_p+\tau(XS_x-2M^2Q^2)-2M\cos\phi_k\sqrt{\lambda_{\tau}\lambda}\bigr),
\nonumber\\&&
\end{eqnarray}
where
\begin{equation}
\lambda_{\tau}=(\tau-\tau_{min})(\tau_{max}-\tau),\quad \lambda=SXQ^2-M^2Q^4-m^2\lambda_Q.
\label{032}\end{equation}

The scalar products of $p_h$ (\ref{009}) are expressed via coefficients
$a^1$,
$a^2$,
$b$,
$a^k$ and
$b^k$:
\begin{eqnarray}\label{a065}
2Ma^1&=&SE_h-(SS_x+2M^2Q^2)p_l\lambda_Q^{-1/2},
\nonumber\\
2Ma^2&=&XE_h-(XS_x-2M^2Q^2)p_l\lambda_Q^{-1/2},
\nonumber\\
b&=&-p_t\sqrt{\lambda/\lambda_Q},
\nonumber\\
2Ma^k&=&E_h-p_l(S_x-2M^2\tau)\lambda_Q^{-1/2},
\nonumber\\
b^k&=&-Mp_t\sqrt{\lambda_{\tau}/\lambda_Q}.
\end{eqnarray}
 The  quantities $E_h$, $P_l$ and $P_t$ are invariants
\begin{eqnarray}\label{a070}
E_h&=&z\nu =\frac{zS_x}{2M},
\nonumber\\
\frac{p_l\sqrt{\lambda_Q}}{M}&=&t-M_h^2+Q^2+2\nu E_h,
\nonumber\\
p_t^2&=&E_h^2-p_l^2-M_h^2.
\end{eqnarray}
  In the rest
 frame they
 take sense of energy,
 longitudinal and transversal momenta of final hadron.


\begin {thebibliography}{99}
%
\bibitem {Poli}
H.Georgi, H.D.Politzer, Phys. Rev. Lett. 40 (1978) 3.
\bibitem {BSh}
      D.Yu.Bardin, N.M.Shumeiko, Nucl. Phys.  B127 (1977) 242.
 \bibitem {ShS}
A.V.Soroko, N.M.Shumeiko, Yad.Fiz. 49 (1989) 1348.
\bibitem {EMC}
EMC Collab., J. Ashman et al, Z. Phys. C52 (1991) 361.
\bibitem {Muld1}
P.J. Mulders, R.D. Tangerman, Nucl. Phys. B461 (1996) 197,
Erratum -- ibid. B484 (1996) 538.
\bibitem {Khoze}
A. Brandenburg, V.V. Khoze, D. Muller,
Phys. Lett. B347 (1995) 413.
\bibitem{Karo}
K.A.Oganessyan, H.R.Avakian, N.Bianchi, P.Di Nezza Eu. Phys. J.
C5 (1998) 681.
\bibitem {POLRAD20}
I.Akushevich, A.Ilyichev, N.Shumeiko, A.Soroko, A.Tolkachev,
 Comp. Phys. Comm. 104 (1997) 201.
\bibitem {AISh}
I.V.Akushevich, A.N.Ilyichev and N.M.Shumeiko, J. Phys.
G24 (1998) 1995.
\bibitem {Ak}
I.Akushevich, hep-ph/9808309, {\it to be published in Eu. Phys. J. C}
\bibitem{hagiwara}
K. Hagiwara, K. Hikasa, N. Kai
 Phys. Rev. D27 (1983) 84.
\bibitem{pak}
N.K.Pak, Ann. Phys. 104 (1977) 54.
\bibitem {JM}
R. Jakob, P.J. Mulders, J. Rodrigues,
 Nucl. Phys. A626 (1997) 937.
\bibitem {Sh}
      N.M.Shumeiko, Sov. J. Nucl. Phys. 29(1979)807.
\bibitem {ASh}
  I.V.Akushevich and N.M.Shumeiko,
  J. Phys. G20 (1994) 513.
\bibitem {delvac}
H. Burkhardt, B. Pietrzyk,
Phys. Lett. B356 (1995) 398.
\bibitem {Manuella}
HERMES Collab. K. Ackerstaff et al., Phys. Rev. Lett.
58 (1998) 5519.
\bibitem {HERMES}
HERMES. Technical design report, 1993.
\bibitem {EMCazim}
EMC Collab., M. Arneodo et al, Z. Phys. C34 (1987) 277.
\bibitem{Aubert}
EMC Collab., J.J.~Aubert {\it et al.}
Phys. Lett. {160B} (1985) 417.
\bibitem{Kramer}
J. Binnewies, B.A. Kniehl, G. Kramer, Z. Phys. C65 (1995) 471.

\end {thebibliography}
\end{document}